# Insights on magnon topology and valley-polarization in 2D bilayer quantum magnets

Doried Ghader

College of Engineering and Technology, American University of the Middle East, Egaila, Kuwait

**Abstract.** The rich and unconventional physics in layered 2D magnets can open new avenues for topological magnonics and magnon valleytronics. In particular, two-dimensional (2D) bilayer quantum magnets are gaining increasing attention due to their intriguing stacking-dependent magnetism, controllable ground states, and topological excitations induced by magnetic spin-orbit couplings (SOCs). Despite the substantial research on these materials, their topological features remain widely unexplored to date. The present study comprehensively investigates the magnon topology and magnon valley-polarization in honeycomb bilayers with collinear magnetic order. We elucidate the separate and combined effects of the SOC, magnetic ground-states, stacking order, and inversion symmetry breaking on the topological phases, magnon valley transport, and the Hall and Nernst effects. The comprehensive analysis suggests clues to determine the SOC's nature and predicts unconventional Hall and Nernst conductivities in topologically trivial phases. We further report on novel bandgap closures in layered antiferromagnets and detail their topological implications. We believe the present study provides important insights into the fundamental physics and technological potentials of topological 2D magnons.

## I. Introduction

The recent experimental realization of 2D magnetic materials[1–3] attracted exceptional interest[4–15] and opened challenging questions related to their fundamental magnetic interactions. In 2D quantum magnets, magnetic anisotropy is crucial to overcome thermal fluctuations and stabilize the magnetic order[16–20]. Recent theoretical and experimental studies promoted the Kitaev interaction[21] induced by the quantum SOC as a candidate to explain the magnetic anisotropy in $CrI_3$, $CrGeTe_3$, and $CrBr_3$ 2D monolayers[22–27]. In parallel, another type of anisotropic SOC, namely the Dzyaloshinskii-Moriya interaction (DMI), received substantial attention because of its potential to explain experimental observations in 2D magnets[26,28–39].

The interest in 2D magnetic materials naturally extends to their magnetic excitations known as spin waves or magnons[25,26,34,40–43]. SOCs in 2D magnets induce topological magnons[27,30–32,34,35,37–40,44–56] with exotic features and great potentials for practical applications. Topological 2D magnets display a finite magnon Hall conductivity that can be harnessed in magnonic nanodevices. Equally important are their topologically protected edge and domain-wall modes, which are robust against structural or magnetic disorders and can form ideal waveguides (or quantum wires) for long-range and coherent magnon transport. Honeycomb $CrI_3$ ferromagnet, for example, represents a model platform to investigate 2D magnetic interactions and topological excitations. Recent studies on monolayer $CrI_3$[25–27] reported large Kitaev couplings and topological magnonic bands with Chern numbers $\mathcal{C} = \pm 1$ in the Heisenberg-Kitaev model[27]. The presence of DMI is also promoted in $CrI_3$ [24,26,34], leaving the question on the nature of the anisotropic interaction in $CrI_3$ and other 2D magnets open for further discoveries.

Stacking layers of 2D ferromagnets is a systematic way to realize new topological quantum materials with richer topology compared to monolayers. Research on layered 2D ferromagnets is exceptionally active due to their intriguing stacking-dependent magnetic behavior and controllable ground states[57–62]. $CrI_3$ bilayers were extensively studied in this context. Their stacking-dependent interlayer coupling leads to ferromagnetic (FM) and layered antiferromagnetic (LAFM) spin configurations in AB and monoclinic stackings, respectively. Similar stacking–dependent magnetism was recently reported in $CrBr_3$ bilayers[63,64], with LAFM state in the $R_{33}$ stacking configuration. AB and monoclinic $CrBr_3$ bilayers, however, favor FM interlayer coupling. Controlled switching between LAFM and FM ground states is also possible in these quantum materials by various experimental techniques[7,60,61,65,66]. For illustration, the FM and LAFM ground states are presented schematically in Figs.1a and 1b, respectively. The monoclinic and $R_{33}$ stackings are illustrated in Supplementary Figs.1a and 1b, respectively.

The SOCs, stacking-dependent magnetism, and controllable magnetic ground states in 2D bilayer magnets constitute novel degrees of freedom that can induce a rich and controllable magnon topology. Advances in manipulating the magnon Hall conductivity, edge magnons, and valley transport in topological 2D bilayer magnets can lead to a new generation of magnonic and spintronic devices[67–69]. Magnon topology in AB-stacked honeycomb FM and LAFM bilayers was analyzed in reference[31], adopting a Heisenberg model with DMI. In both configurations, the magnon spectrum was found topological (Zeeman effect is necessary in the LAFM case) with a single phase. In a recent study[39], we have extended the analysis on AB-stacked honeycomb FM bilayers, including layer-dependent electrostatic doping[66] (ED). Five distinct topological phases were predicted as a result of the interplay between the model parameters. Moreover, recent theoretical studies[70,71] initiated the discussion on topological magnon valley transport in AB-stacked FM and LAFM bilayers. Nevertheless, a comprehensive investigation on the magnon topology and valley transport in honeycomb collinear FM and LAFM bilayers is still missing.

An essential step towards low-dimensional topological magnonics is to understand the implications of quantum interactions, symmetry breaking scenarios, magnetic ground states, and stacking orders on the magnonic topology and transport in 2D layered magnets. The present study serves this ambitious goal by analyzing the magnon topology in an extended group of honeycomb bilayer models with collinear magnetic order. The models are generated from all relevant combinations of Kitaev and/or DM interactions, inversion symmetry breaking terms, collinear FM and LAFM states, and standard stacking orders (AB, monoclinic, and $R_{33}$, in particular). The comprehensive study reports several unconventional findings and elucidates the magnon topology's strong dependence on the stacking order, magnetic ground state, and SOC type. Finite Hall and Nernst conductivities are proved possible in FM and LAFM phases with zero Chern numbers. LAFM bilayer models reveal anomalous bandgap closures at unconventional nodal points (UNPs) formed away from the high symmetry points. These play a primary role in shaping the topological phase diagrams (TPDs) for LAFM bilayers. We further discuss opportunities to realize magnon valley transport in all models and suggest routes to determine the SOC's nature in several models based on the Hall and Nernst conductivities' profiles. The study aligns with current efforts to manipulate topological magnons in quantum 2D magnets towards their implementation in low-dimensional spintronics and magnonics.

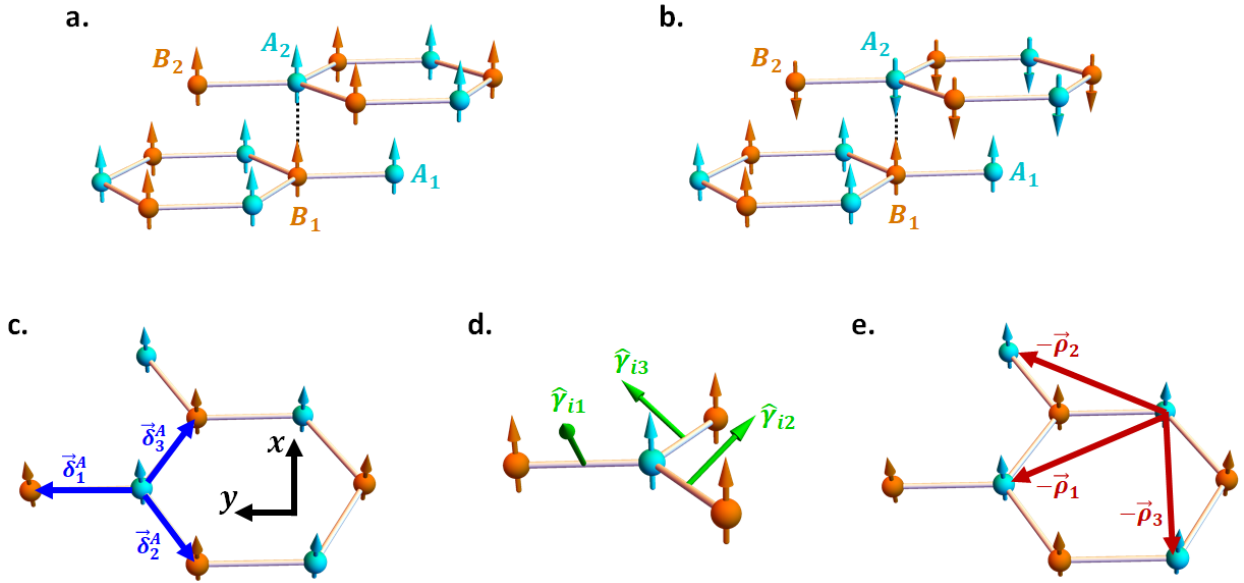

**Figure 1:** Schematic representation of FM (a) and LAFM (b) bilayers, respectively. (c) and (e) respectively illustrate the position vectors for an A site's nearest and next-nearest neighbors. Figure (d) presents the Kitaev unit vectors for an A site (green vectors).

Before proceeding, we will set conventions and definitions adopted throughout the paper. The strength of the Kitaev interaction, DMI, and inversion symmetry breaking term will be denoted $K$, $D$, and $U$ respectively. The bold notation $\pm \boldsymbol{K}$ is used for the Brillouin zone (BZ) corners (or valleys). We break the inversion symmetry by introducing ED in the FM case and Zeeman effect due to an external magnetic field in the LAFM case. By default, all models include intralayer and interlayer Heisenberg exchange interactions. The short labels for our models are listed in Table 1. The four magnonic bands are termed $\epsilon_1, \dots, \epsilon_4$ in ascending energy order, and the color code for the corresponding band closure manifolds is defined in Table 2. Nonadiabatic paths represented by black-dashed curves are introduced in selected topological phase diagrams (TPDs). These paths help investigate the nonadiabatic evolution of the thermal Hall and Nernst conductivities. The conductivities are temperature-dependent and reach saturation at sufficiently high temperatures. Our figures present the maximum (saturation) values of these conductivities. Of note, the numerical results presented in the study are representative samples of our extensive numerical analysis. All stated conclusions are general and independent of the specific parameters used in the plots.

| Label | Model description |
|---|---|
| $U$ model | $U \neq 0$, $D = 0$, and $K = 0$. |
| $D$ model | $U = 0$, $D \neq 0$, and $K = 0$. |
| $K$ model | $U = 0$, $D = 0$, and $K \neq 0$. |
| $K + D$ model | $U = 0$, $D \neq 0$, and $K \neq 0$. |
| $D + U$ model | $U \neq 0$, $D \neq 0$, and $K = 0$. |
| $K + U$ model | $U \neq 0$, $D = 0$, and $K \neq 0$. |
| $K + U + D$ model | $U \neq 0$, $D \neq 0$, and $K \neq 0$. |

**Table 1.** Short labels for the studied models.

| Manifold color | Manifold Description |
|---|---|
| Green | Closes the gap between $\epsilon_1$ and $\epsilon_2$ at $+\mathbf{K}$. |
| Red | Closes the gap between $\epsilon_2$ and $\epsilon_3$ at $+\mathbf{K}$. |
| Purple | Closes the gap between $\epsilon_3$ and $\epsilon_4$ at $+\mathbf{K}$. |
| Blue | Closes the gap between $\epsilon_3$ and $\epsilon_4$ at $-\mathbf{K}$. |
| Orange | Closes the gap between $\epsilon_2$ and $\epsilon_3$ at $-\mathbf{K}$. |
| Cyan | Closes the gap between $\epsilon_2$ and $\epsilon_3$ at unconventional nodal points away from the high symmetry points. |

**Table 2.** Color code for the band closure manifolds in the TPDs.

## II. Results

**Magnonic Hamiltonian**

The real-space Hamiltonian describing the various interaction in the collinear FM or LAFM honeycomb bilayer can be expressed as

$$\mathcal{H} = -J \sum_{l,i,j} \vec{S}_{li} \cdot \vec{S}_{lj} - K \sum_{l,i,j} (\vec{S}_{li} \cdot \hat{\gamma}_{ij})(\vec{S}_{lj} \cdot \hat{\gamma}_{ij}) - J_\perp \sum_{i,j} \vec{S}_{1i} \cdot \vec{S}_{2j} + \sum_{l,m,n} D_{mn} \vec{S}_{lm} \cdot \vec{S}_{ln}^D$$
$$- \sum_{l,i} U_l \hat{z} \cdot \vec{S}_{li} - \mathcal{A} \sum_{l,i} (\vec{S}_{li} \cdot \hat{z})^2$$

(1)

The coefficients $J$, $K$, $J_\perp$, and $D_{mn}$ determine the strength of the intralayer exchange, Kitaev interaction, interlayer exchange, and DMI, respectively. $\mathcal{A}$ stands for the easy axis anisotropy parameter. The inversion symmetry breaking parameter $U_l$ will be defined shortly. The vector $\vec{S}_{li}$ denotes the spin on a site $i$ in layer $l$ ($l = 1, 2$ for the bottom and top layers, respectively). In the FM state, the spins in both layers can be expressed as $\vec{S} = (S_x, S_y, S_z)$. The LAFM configuration follows from the FM case by a $\pi$ rotation of the top layer along the y-axis. Consequently, the Hamiltonian in Eq.1 stays valid for the LAFM bilayers, with the spins in layer 2 expressed as $\vec{S} = (-S_x, S_y, -S_z)$.

The first, second, and third terms correspond to the nearest neighbor (NN) intralayer exchange, intralayer Kitaev, and interlayer exchange interactions, respectively. The indices $i$ and $j$ in these three terms are summed over the NN. The Kitaev contribution in Eq.1 deserves further illustration as follows. For an $A$ spin at site $i$, the Kitaev unit vectors are $\hat{\gamma}_{i1} = \left(-\frac{\sqrt{2}}{\sqrt{3}}, 0, \frac{1}{\sqrt{3}}\right)$, $\hat{\gamma}_{i2} = \left(\frac{1}{\sqrt{6}}, -\frac{1}{\sqrt{2}}, \frac{1}{\sqrt{3}}\right)$, and $\hat{\gamma}_{i3} = \left(\frac{1}{\sqrt{6}}, \frac{1}{\sqrt{2}}, \frac{1}{\sqrt{3}}\right)$, associated with the three NN links $\vec{\delta}_1^A = a\left(0, \frac{1}{\sqrt{3}}, 0\right)$, $\vec{\delta}_2^A = a\left(-\frac{1}{2}, -\frac{1}{2\sqrt{3}}, 0\right)$, and $\vec{\delta}_3^A = a\left(\frac{1}{2}, -\frac{1}{2\sqrt{3}}, 0\right)$, respectively. The NN links and Kitaev unit vectors are illustrated in Figs.1c and 1d, respectively. For a site on the B sublattice, the link vectors flip sign ($\vec{\delta}_i^B = -\vec{\delta}_i^A$) while the Kitaev unit vectors remain invariant.

The fourth term in $\mathcal{H}$ corresponds to the DM interaction, and the indices $m$ and $n$ are summed over the six next nearest neighbors. These are situated at relative position vectors, $\vec{\rho}_1 = a\left(\frac{1}{2}, -\frac{\sqrt{3}}{2}, 0\right)$, $\vec{\rho}_2 = a\left(-\frac{1}{2}, -\frac{\sqrt{3}}{2}, 0\right)$, $\vec{\rho}_3 = a(1, 0, 0)$, $\vec{\rho}_4 = -\vec{\rho}_1$, $\vec{\rho}_5 = -\vec{\rho}_2$, and $\vec{\rho}_6 = -\vec{\rho}_3$ (see Fig. 1e). We have introduced the vector $\vec{S}_{ln}^D = (S_{lny}, -S_{lnx}, 0)$ which transforms the DMI contribution to a scalar-product rather than a cross-product[39]. The coefficient $D_{mn} = \pm D$, with the sign determined in the conventional way from the local geometry of the honeycomb lattice[30].

The last term in $\mathcal{H}$ accounts for the easy-axis magnetic anisotropy, while the fifth term is introduced to break the inversion symmetry. For LAFM bilayers, the fifth term is induced by an external magnetic field $\vec{h} = U\hat{z}$, setting $U_l = \pm U$ for layers 1 and 2, respectively. For the FM bilayer case, we adopt the approach suggested in reference[70] to break the inversion symmetry via layer-dependent ED. In particular, the ED technique can efficiently tune the localized moments in each layer[66,72], generating opposite potentials ($\pm U$) when the two monolayers are doped equally with opposite charges. An alternative approach is to induce layer-dependent magneto-crystalline anisotropy in the FM bilayer. This can be achieved when the FM bilayer is sandwiched between a substrate and a superstrate composed of different materials. The sandwiched structure ensures a distinct atomic environment for each layer and allows independent manipulation of the monolayers' magnetic properties, including the magneto-crystalline anisotropy

In Supplementary Notes 1 and 2, we develop the semi-classical linear spin-wave approach[73–78] and derive the FM and LAFM momentum ($\vec{p}$) space Hamiltonians for bilayers stacked in the AB, monoclinic, and $R_{33}$ configurations. The Holstein-Primakov method yields identical results. To our knowledge, the semi-classical approach is not yet developed for FM and LAFM bilayers with Kitaev interaction, which made us prefer it over the Holstein-Primakov formalism. The Hamiltonians are derived as $8 \times 8$ matrices and produce four physical bands. We will use the notation $[\mathcal{C}_4, \mathcal{C}_3, \mathcal{C}_2, \mathcal{C}_1]$ to group the Chern numbers of the bands $[\epsilon_4, \epsilon_3, \epsilon_2, \epsilon_1]$, respectively. Details on the Berry curvatures and Chern numbers calculation are presented in the Methods section.

Before proceeding, we fix the value of the easy-axis anisotropy parameter $\mathcal{A}$. This parameter is essential to stabilize the magnetic order in real 2D magnets. Nevertheless, as long as magnons are concerned, the effect of $\mathcal{A}$ is a uniform upward shift in the magnon spectra. $\mathcal{A}$ consequently lifts the degeneracy between positive and negative solutions, which is imperative for topological calculations. The specific value of $\mathcal{A}$ is, however, irrelevant to the topological results. We will fix the easy axis anisotropy at $\mathcal{A} = 0.25J$ in the numerical calculations, chosen arbitrarily as per the above argument.

**Topological analysis in AB-stacked FM bilayers**

The $D$ model for AB-stacked FM bilayers was studied in reference[31], revealing a single topological phase with Chern numbers $[0, -2, 0, 2]$. The rich magnon topology in the FM $D + U$ model was discussed in our recent work[39]. The model hosts five topological phases with Chern numbers $[0, -2, 0, 2]$, $[0, -2, 1, 1]$, $[-1, -1, 1, 1]$, $[-1, 1, -1, 1]$, and $[0, 0, -1, 1]$. The corresponding topological phase transitions are accompanied by bandgaps closure at the $\pm K$ BZ corners exclusively. The $U$ model for these bilayers was extensively analyzed in references[70,71], reporting valley-polarized magnons in a remarkable analogy with electronic 2D materials. These three models are hence excluded here.

We start with the $K$ model for AB-stacked FM bilayers. The magnon spectrum in the present case is gapped and topological for any nonzero value of $K$, with Chern numbers $[0,-2,0,2]$. This is the same topological phase encountered in the $D$ model[31]. An illustrative example of the gapped 4-band magnon spectrum is presented in Fig.2a for $J_\perp = 0.3J$ and $K = 2J$. The bands are plotted along the high symmetry axes $K\Gamma$, $\Gamma M$, and $MK$ of the conventional honeycomb BZ (not shown). Including the DMI ($K + D$ model) augments the bandgaps without affecting the above conclusions regarding the magnon topology.

In the $K + U$ model, the magnon band spectrum is gapped except at specific 3D manifolds in the $(K, U, J_\perp)$ parametric space. These manifolds close the gaps at the BZ corners ($\pm K$) and lead to topological phase transitions. To illustrate, the TPD is presented in Fig.2b over a sufficiently large range of the normalized Kitaev interaction ($K/J$) and ED potentials ($U/JS$) with $J_\perp = 0.2J$. $S$ denotes the z-component of the spin (Supplementary Note 1). The color code for the gap closure manifolds is defined in Table 2. The manifolds divide the parametric space into five topological regions (or phases), denoted I, …,V. Their respective Chern numbers are $[0,-2,0,2]$, $[0,-2,1,1]$, $[-1,-1,1,1]$, $[-1,1,-1,1]$, and $[0,0,-1,1]$. The topological phases in the $K + U$ model match those of the $D + U$ model investigated in our previous study[39]. Consequently, the DM and Kitaev interactions in AB-stacked FM bilayers cannot be easily differentiated in terms of their topological consequences. Region VI, however, is gapped but topologically trivial with zero Chern numbers. Interestingly, the Berry curvatures in VI do not vanish and are peaked at the BZ corners (Figs.2c and 2d) with opposite signs near $\pm K$. This sign behavior is absent in the topological phases. The Berry curvatures' profile in VI is similar (but not identical) to the $U$ model, which was proved to support topological valley transport [70,71]. We will shortly illustrate the consequences of the nonzero Berry curvatures on the standard and valley Hall conductivities in VI.

The rich topology in the $K + U$ model motivates the analysis of the topological conductivities' nonadiabatic evolution. The black-dashed path in Fig.2b is defined parametrically as $\left\{U = 0.4\sqrt{\cos(u)}, K = 2.8\sin(u), \frac{5}{34}\pi < u < \frac{1}{2}\pi\right\}$. Figs.2e and 2f show the (saturated) magnon Hall ($\kappa_{xy}$) and Nernst ($\alpha_{xy}$) conductivities evolving nonadiabatically with preserved signs along this path. This is a general result and not related to the chosen path. Reversing the signs of the conductivities by topological phase transitions is hence not possible in the present model. The plots further demonstrate discontinuities in the $\kappa_{xy}$ and $\alpha_{xy}$ profiles with abrupt variations accompanying the topological phase transitions. An additional example is presented in Supplementary Fig. 1d along a nonadiabatic path with weak ED (Fig. 1c). The region with weak ED might be particularly interesting for experimental studies. The Hall conductivity in this region is found to present an exotic profile, with significantly abrupt jumps and asymptotic-like behavior near the boundaries. Importantly, $\kappa_{xy}$ and $\alpha_{xy}$ do not necessarily vanish in the trivial phase VI (Fig.2e, Fig.2f, and Supplementary Fig. 1d). For sizable Kitaev interactions, the absolute values of the Berry curvature's peaks differ at $\pm K$ (Fig.2c), resulting in net Hall and Nernst conductivities. For weaker (but nonnegligible) Kitaev interactions, the peaks' absolute values are equal at $\pm K$ (Fig.2d), in analogy with the $U$ model[70,71]. Consequently, the standard conductivities $\kappa_{xy}$ and $\alpha_{xy}$ vanish, while the valley Hall and Nernst conductivities become finite[70].

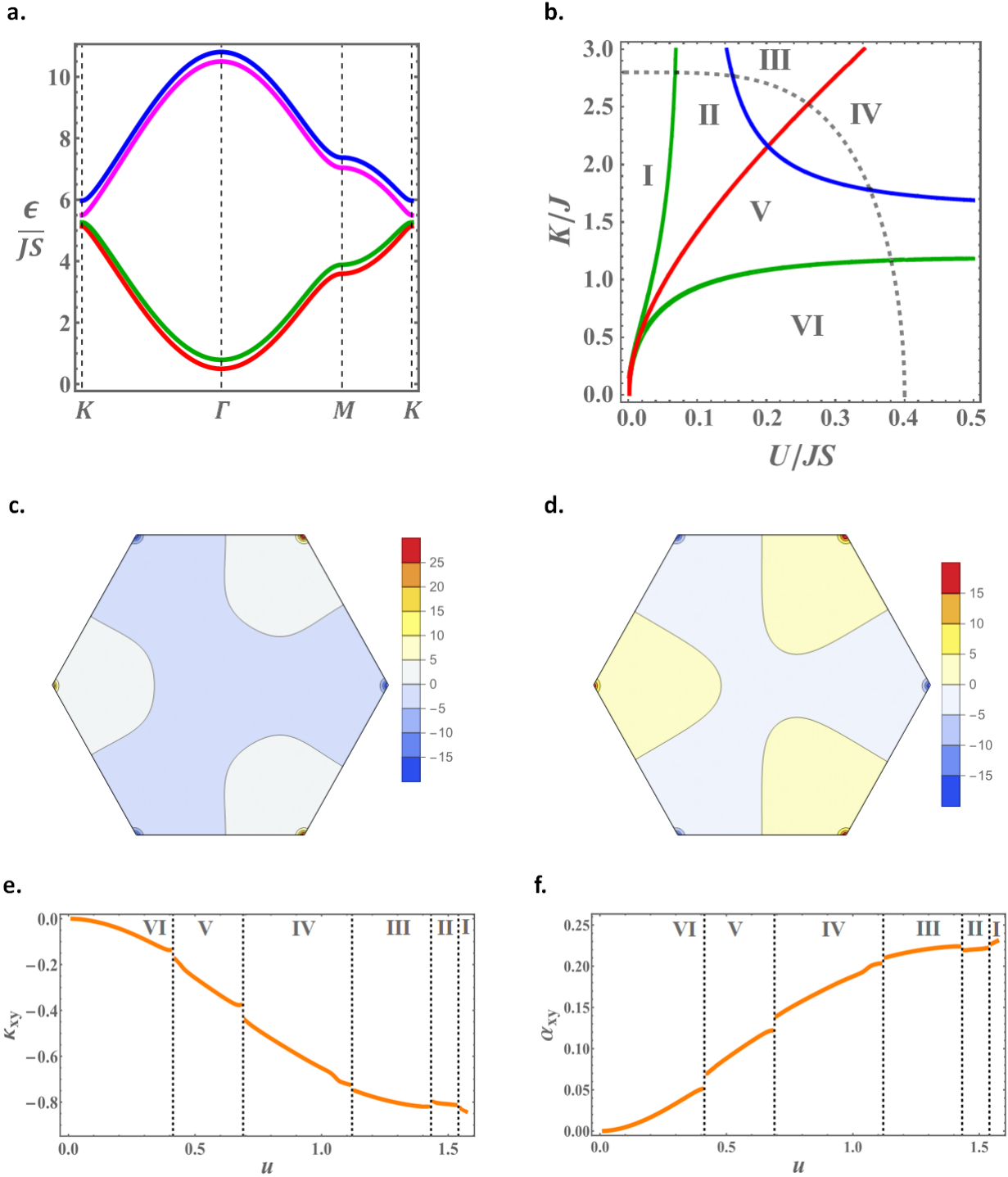

**Figure 2:** (a) The $K$ model gapped 4-band magnon spectrum plotted along high-symmetry axes for parametric values $K = 2J$ and $J_\perp = 0.3J$. (b) TPD for the $K + U$ model with $J_\perp = 0.2J$. The black-dashed curve represents a nonadiabatic path in the parametric space. (c) and (d) are the Berry curvatures for band $\epsilon_4$ in the $K + U$ model's trivial phase with $(U = 0.25JS, K = 0.58J)$ and $(U = 0.25JS, K = 0.2J)$, respectively. The rest of the bands show similar behavior. (e) and (f) illustrate the nonadiabatic evolution of the Hall and Nernst conductivities, respectively. These are plotted along the nonadiabatic path in (b).

In the most general $K + U + D$ model for AB-stacked FM bilayers, the parametric space is 4D, generated by the Hamiltonian parameters $(K, U, J_\perp, D)$. The DMI added to the Kitaev interaction is found to reconstruct the gaps closure manifolds without inducing new topological phases compared to the previous $K + U$ model. Nevertheless, the DMI wipes out the non-topological phase rendering the model topological for any parametric combination away from the gap closure manifolds. Moreover, the behavior of $\kappa_{xy}$ and $\alpha_{xy}$ in the $K + U + D$ model does not show any relevant difference compared to the $K + U$ model: the nonadiabatic evolution preserves the signs and induces abrupt jumps across the gap closure manifolds. The $K + U + D$ model TPD and conductivities are presented in Supplementary Fig.1.

Before closing this section, we point out the important implications of the magnon spectrum's nonreciprocity ($\epsilon_i(\vec{p}) \neq -\epsilon_i(-\vec{p})$) in models with both ED and SOCs. The independent gap closures at $\pm\boldsymbol{K}$ (Fig.2b) are a direct consequence of this bands' nonreciprocity. Moreover, in the $K + U$ model, the bands' nonreciprocity plays an essential role in realizing the Hall and Nernst conductivities in the trivial phase VI. Bands' nonreciprocity holds in all models with sizable ED and SOCs, regardless of the magnetic ground state and stacking order.

**Topological analysis in AB-stacked LAFM bilayers**

In the AB-stacked LAFM configuration, a perpendicular external magnetic field is indispensable to lift the degeneracy and induce bandgaps. Models with $U = 0$ are hence irrelevant to the present study. Like the FM case, the bandgaps in AB-stacked LAFM models with DMI and/or Kitaev interactions can close at $\pm\boldsymbol{K}$, resulting in numerous topological phases. Nevertheless, the magnon spectra in LAFM bilayers have an essential aspect that is absent in FM bilayers. Namely, they display UNPs that close the bandgaps away from $\pm\boldsymbol{K}$, specifically between $\epsilon_2$ and $\epsilon_3$ bands. These play a crucial role and enrich the magnon topology in LAFM bilayers.

Reference[70] discussed the $U$ model for AB-stacked LAFM bilayers. Here, we complement the study and highlight the role of the UNPs in this model. The magnon spectrum is gapped at $\pm\boldsymbol{K}$ for any nonzero value of $U$. Nevertheless, the interplay between $J_\perp$ and $U$ can close the gap between $\epsilon_2$ and $\epsilon_3$ at the UNPs. Examples of the UNPs and the corresponding bandgap closures are presented in the left panel of Fig.3a. In the right panel, we plot bands $\epsilon_2$ and $\epsilon_3$ over the shaded region centered at the $+\boldsymbol{K}$ corner of the BZ. The bandgap between $\epsilon_2$ and $\epsilon_3$ reaches minimal values along the contour highlighted in red. The bandgap closes at the UNPs near the intersection between the red contour and the high symmetry axes $\boldsymbol{K\Gamma}$ and $\boldsymbol{MK}$ (left panel of Fig.3a). This discussion is valid for all LAFM models presenting UNPs.

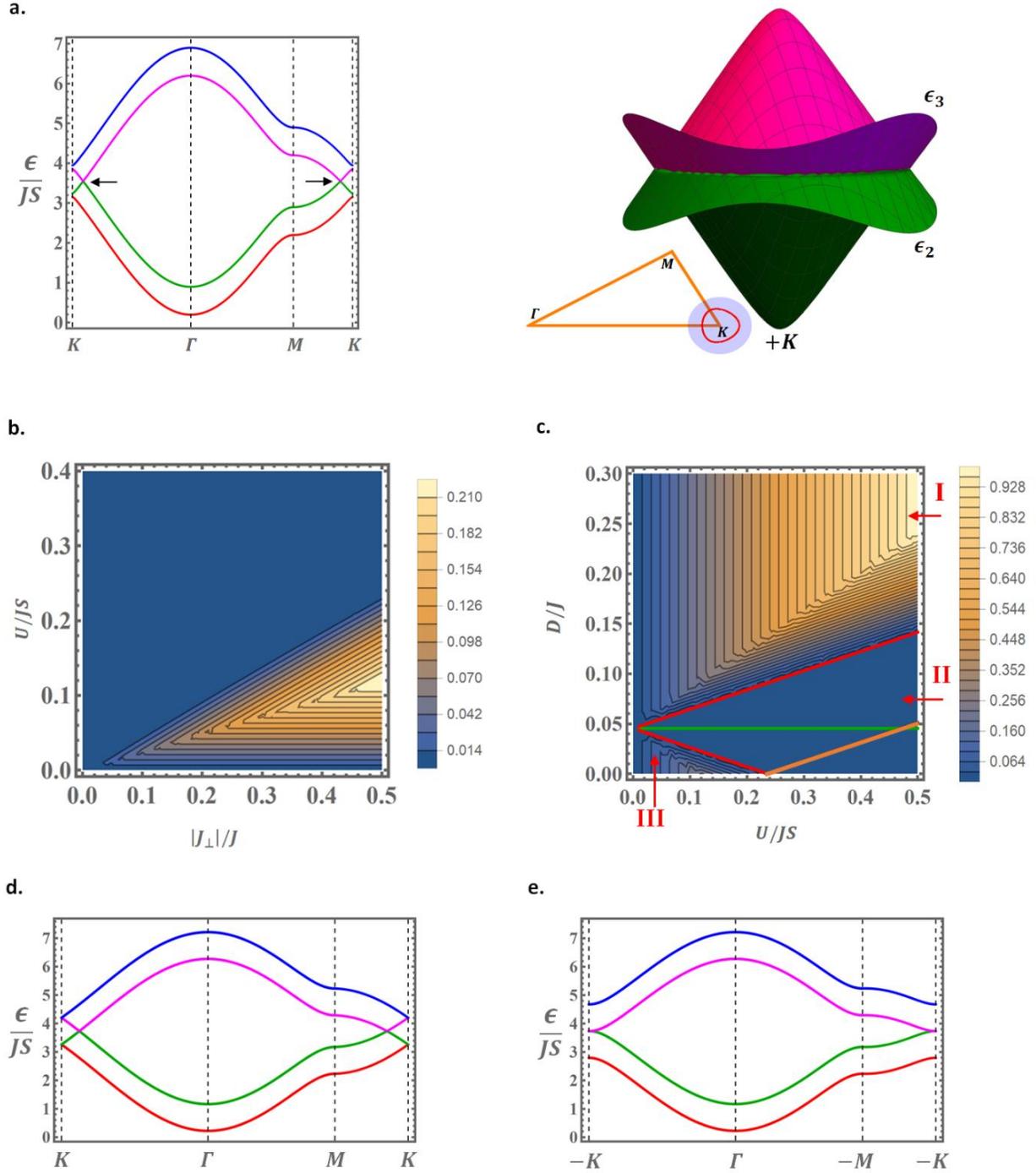

**Figure 3:** (a) *Left panel:* UNPs (highlighted with black arrows) in the $U$ model for AB-stacked LAFM bilayers with $J_\perp = -0.1J$ and $U = 0.35JS$. *Right panel:* Plots of $\epsilon_2$ and $\epsilon_3$ over the blue shaded disk (radius $0.7/a$) centered at $+K$. The UNPs are located on the red contour where the bandgaps are tiny. (b) Contour plot for the minimal gap between $\epsilon_2$ and $\epsilon_3$ in the $U$ model, plotted in the $(U, J_\perp)$ plane. The interlayer interaction ($J_\perp$) is presented in absolute values. (c) Contour plot for the minimal bandgap in the $D+U$ model for $J_\perp = -0.5J$. Three phases are highlighted: topological phase I, ungapped phase II, and gapped but trivial phase III. The colored manifolds close the bandgaps at $\pm K$ according to the color code in Table 2. (d) and (e) illustrate different bandgap closure scenarios with the parameters corresponding to the intersection point between the orange, pink and green manifold in (c).

Fig.3b plots the minimal bandgap in the $U$ model as a function of the normalized external field ($U/JS$) and interlayer interaction ($J_\perp/J$). The contour plot reveals a gapped phase in weakly coupled LAFM bilayers. The Berry curvatures are nonzero in the gapped phase, with antisymmetric peaks near the $\pm K$ valleys. Hence, the standard Hall and Nernst conductivities vanish, but valley-polarized magnons can be generated in the gapped region, as pointed out in reference[70].

The $D + U$ model in AB-stacked LAFM bilayers was first studied in reference[31], excluding the important consequences of the UNPs. Fig.3c presents the minimal bandgap's contour plot in the $DU$ space for an exaggerated value of the interlayer exchange, $J_\perp = -0.5J$. The figure equally serves as a TPD. The TPD preserves its main features for more realistic (weaker) $J_\perp$ in van der Waals (vdWs) magnets. The numerical results reveal a topological phase I, an ungapped phase II with gap closures at UNPs, and a gapped but trivial phase III with zero Chern numbers. Phase III disappears for weaker interlayer exchange ($|J_\perp| < 0.2J$ as a rough estimation).

The $\pm K$ gap closure manifolds still exist (Fig.3c), notably in phase II, but do not contribute to the magnon topology. The green manifold in Fig.3c coincides with the pink one, simultaneously closing the ($\epsilon_1$, $\epsilon_2$) and ($\epsilon_3$, $\epsilon_4$) gaps at $+K$ (see the color code in Table 2). Illustrative examples of the gap closure scenarios are presented in Figs.3d and 3e at the intersection between the green, pink, and orange manifolds.

The Berry curvatures (not shown) are nonzero in the trivial phase III and have opposite-sign peaks near $\pm K$. Nevertheless, the absolute values of the peaks are significantly different at $\pm K$. Hence, the trivial phase can support standard Hall and Nernst conductivities, but magnon valley transport seems unlikely in this phase.

Replacing the DMI by Kitaev interaction enriches the magnon topology drastically. Unlike the $D + U$ model, both conventional and unconventional gap closure manifolds conspire to shape the topology in the $K + U$ model. The TPD is presented in Fig.4a for the $KU$ slice of the parametric space corresponding to $J_\perp = -0.3J$. The structure of the TPD is preserved for smaller/larger values of $J_\perp$. The UNPs are encountered only along the cyan manifolds forming minor regions in the parametric space. The ensemble of band closure manifolds divides the $KU$ space into seven different gapped phases labeled I,…,VII. Away from the cyan manifolds, the gaps between $\epsilon_2$ and $\epsilon_3$ are sizable in phases I and VII, and tiny elsewhere. The Berry curvature calculation proves that phases I,…,VI are topological with characteristic Chern numbers $[-1,1,1-1]$, $[-1,2,0,-1]$, $[-1,-1,3,-1]$, $[0,-2,3,-1]$, $[0,-2,2,0]$, and $[0,1,-1,0]$. Phase VII, however, is gapped and trivial. Similar to the previously encountered trivial phases, region VII supports standard and valley Hall conductivities for sizeable and weak (but not negligible) DMI, respectively.

The drastic difference between the magnon topologies in the $D + U$ model (1 topological phase) and the $K + U$ model (6 topological phases) is remarkable. The magnon topology in the $K + U$ model is tunable via experimentally controllable parameters, like the external magnetic field and interlayer interactions[60]. While topological phase transitions are unlikely in the $D + U$ model, they can be experimentally realized in the $K + U$ counterpart. This can help determine the nature of the SOC in AB-stacked 2D magnets with LAFM ground state.

Furthermore, the rich magnon topology in the $K + U$ model induces novelties in the profiles of $\kappa_{xy}$ and $\alpha_{xy}$. These are plotted in Figs.4c and 4d along the nonadiabatic directional path of Fig.4a, specified by $\left\{U = 0.4\cos^4(u), K = 2\sin(u) + \cos^2(u) + 1, -\frac{\pi}{2} < u < \frac{\pi}{2}\right\}$. The topological phase transitions induce abrupt jumps and can even lead to asymptotic-like behaviors at some boundaries. Experimental observation of any sharp variation in $\kappa_{xy}$ and $\alpha_{xy}$ shall constitute an unambiguous signature of the Kitaev interaction since topological phase transitions are absent in the $D + U$ model. More importantly, the topological phase transitions in the $K + U$ model can reverse the signs of $\kappa_{xy}$ and $\alpha_{xy}$, revealing a unique feature of AB-stacked LAFM Kitaev magnets. The sign reversal is best manifested in the Hall conductivity, showing two asymptotic-like behaviors of opposite signs at the (V-IV) and (IV-III) boundaries. As will be demonstrated shortly, these signatures are absent in samples with relevant values of the DMI. Consequently, experimental observation of conductivity sign reversal eliminates the chances for sizable DMI and uncovers the SOC's exclusive Kitaev nature.

Before proceeding to the $K + U + D$ model, we note that the signs of $\kappa_{xy}$ and $\alpha_{xy}$ in the $D + U$ model's single topological phase can be reversed only by reversing the direction of the external magnetic field[31]. However, the observed sign reversal in the $K + U$ model with preserved magnetic field direction is of fundamental nature and related to the topological phase transitions. This sign reversal shall be detectable in experiments with appropriate control on the external magnetic field's strength and/or the interlayer exchange.

The phase diagram in the $K + U + D$ model for AB-stacked LAFM bilayers is found to be significantly different compared to the $K + U$ model. The DMI enhances the gaps between $\epsilon_2$ and $\epsilon_3$, which can only close in bilayers with tiny DMI and exaggerated interlayer exchange. The reduced number of band closure manifolds decreases the number of topological phases, leaving only 2 of them in bilayers with sizable DMI and/or weak vdWs interlayer exchange. The TPD is presented in Fig.4b for $J_\perp = -0.2J$ and $D = 0.05J$. Similar to the FM case, the $K + U + D$ model does not allow topologically trivial phases. The Chern numbers are $[-1,1,1,-1]$ and $[-1,2,0,-1]$ for phases I and II, respectively. These match the first two topological phases reported in the previous $K + U$ model. Topological phase transitions between I and II cannot reverse the signs of the Hall and Nernst conductivities. Consequently, sign reversal of $\kappa_{xy}$ and $\alpha_{xy}$ is not allowed in AB-stacked LAFM bilayers with relevant DMI and/or weak (vdWs) interlayer exchange. We hence restate our conclusion that sign reversals unambiguously reveal the exclusive Kitaev nature of the SOC.

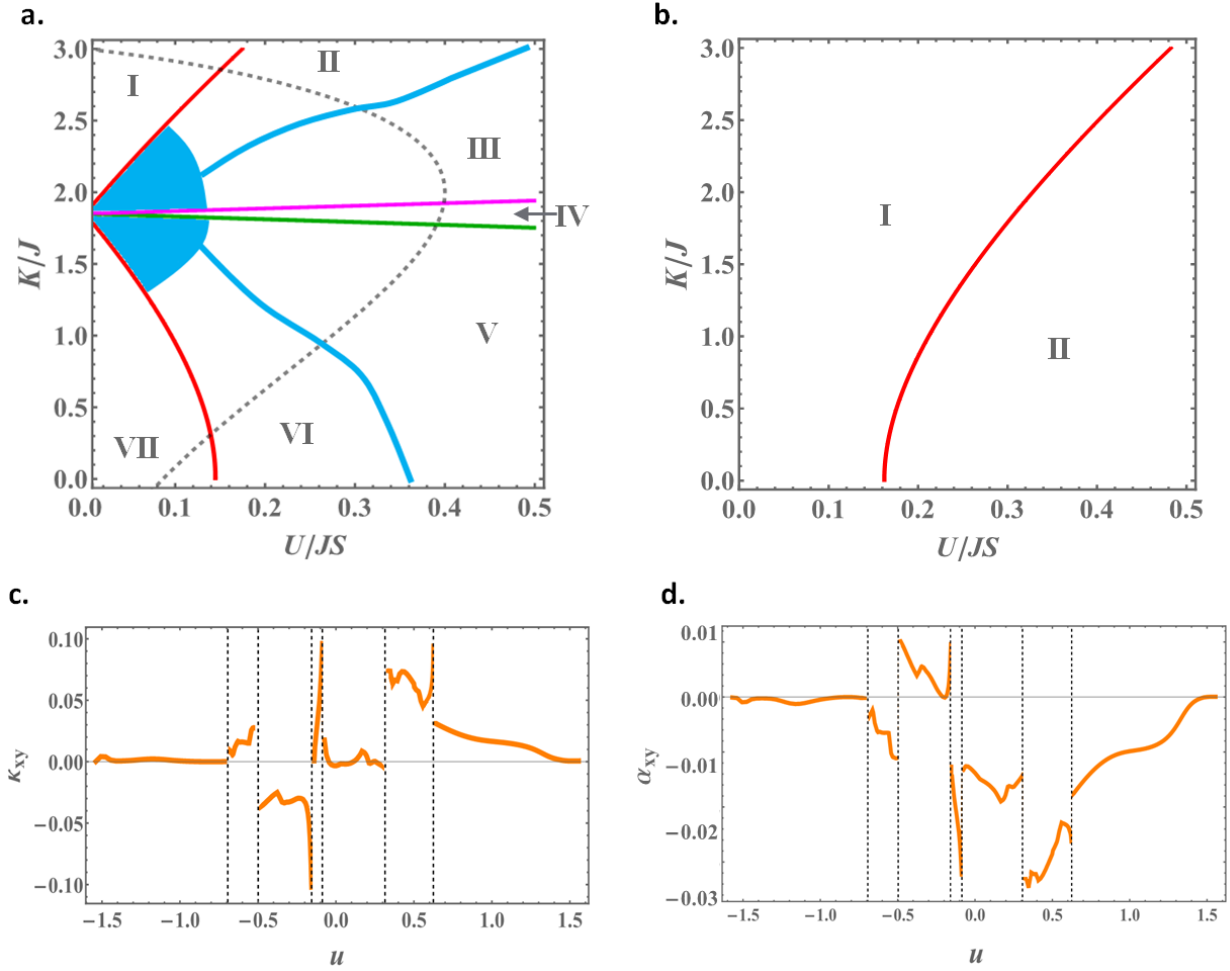

**Figure 4:** (a) TPD in the $K+U$ model for AB-stacked LAFM bilayers with $J_\perp = -0.3J$. (b) TPD for $K+U+D$ bilayers with $J_\perp = -0.2J$ and $D = 0.05J$. (c) and (d) illustrate the nonadiabatic evolution of the $K+U$ Hall and Nernst conductivities, respectively. These are plotted along the nonadiabatic path in (a).

**Topological analysis in monoclinic FM and LAFM bilayers**

The previous section elucidated the significant topological difference between FM and LAFM ground states in AB-stacked bilayers. In this section, the stacking-dependent topology is exposed. We start with monoclinic FM bilayer models. The $U$ model is very similar to the AB-stacked case and supports topological magnon valley transport for any $U$. The $K$, $D$, and $K+D$ models are gapped with a single topological phase $[1,1,-1,-1]$, which is different from the $[0,-2,0,2]$ phase in the AB-stacking. The magnonic Hamiltonian hence evolves nonadiabatically between the AB and the monoclinic stacking. The remaining models ($K + U$, $D + U$, and $K + U + D$) are topologically identical to the $K$ and $D$ models. The rich magnon topology reported in the AB-stacked FM bilayers is indeed absent in the monoclinic case.

We arrive at LAFM monoclinic bilayers with some novel results. Unlike the AB-stacked case, the monoclinic $U$ model is not gapped, and magnon valley transport is not possible. The bandgaps in the $U$ model close at $\pm \boldsymbol{K}$ and/or the UNPs, depending on the value of $J_\perp$ and $U$. The $K$ and $D$ models are topologically trivial, where $\epsilon_1$ and $\epsilon_2$ bands are degenerate throughout the BZ (similarly for $\epsilon_3$ and $\epsilon_4$). Next, the TPD for the $D + U$ model is plotted in Fig.5a with $J_\perp = -0.1J$. Phase I is topological, with Chern numbers $[-1,1,1,-1]$, while phase II is ungapped. The bandgaps close at UNPs in phase II. Phase II can be realized experimentally via ED. This ungapped phase is absent in the $K + U$ and $K + U + D$ models, presenting phase I only. Consequently, the realization of an ungapped phase in LAFM monoclinic bilayers reveals the exclusive DMI nature of the SOC. We also note that topological magnon valley transport is absent in all monoclinic LAFM models.

**Topological analysis in $R_{33}$ stacked FM and LAFM bilayers**

To our knowledge, magnon topology in $R_{33}$ stacked bilayers was not discussed in any previous work. The results for the FM $U$ model are very similar to the AB and monoclinic cases. For FM bilayers with SOC, the $R_{33}$ configuration presents new results that are absent in the AB and monoclinic cases. Fig.5b illustrates the rich $D$ model TPD. Regions I, II, and III are topological with Chern numbers $[1,1,-1,-1]$, $[1,0,0,-1]$, and $[1,-1,1,-1]$, while IV is gapped but trivial. The $K$ model (Fig.5c) hosts four topological phases I, II, III, and IV with Chern numbers $[1,1,-1,-1]$, $[1,0,0,-1]$, $[1,-1,1,-1]$, and $[1,-1,0,0]$. Phase V in the $K$ model is gapped but trivial. These results are substantially different from the $D$ and $K$ models in the AB and monoclinic FM bilayers, presenting a single topological phase. The topologically trivial phases in the $D$ and $K$ models support standard or valley Hall transport, in analogy with the previously discussed models. The $K + D$ model presents a single topological phase with Chern numbers $[1,1,-1,-1]$, while the $D + U$ model hosts two topological phases I and II (Figs.5d) with Chern numbers $[1,1,-1,-1]$ and $[1,-1,1,-1]$, respectively. The $K + U$ model is richer (Figs.5e), with five phases topologically identical to those of the $K$ model. Finally, the most general $K + U + D$ model (Figs.5f) displays three topological phases I, II, and III, with respective Chern numbers $[1,1,-1,-1]$, $[1,0,0,-1]$, and $[1,-1,1,-1]$. The nonadiabatic behaviors of $\kappa_{xy}$ and $\alpha_{xy}$ (not shown) in all $R_{33}$ models with SOCs are comparable to Figs.2e and 2f. The topology of the $K + U$, $D + U$, and $K + U + D$ models can be tuned by $U$, and the significant topological difference between these models might be useful to explore the nature of the SOC. Finally, the topological phase diagrams and related conclusions in $R_{33}$ stacked LAFM models are very similar to their monoclinic counterparts.

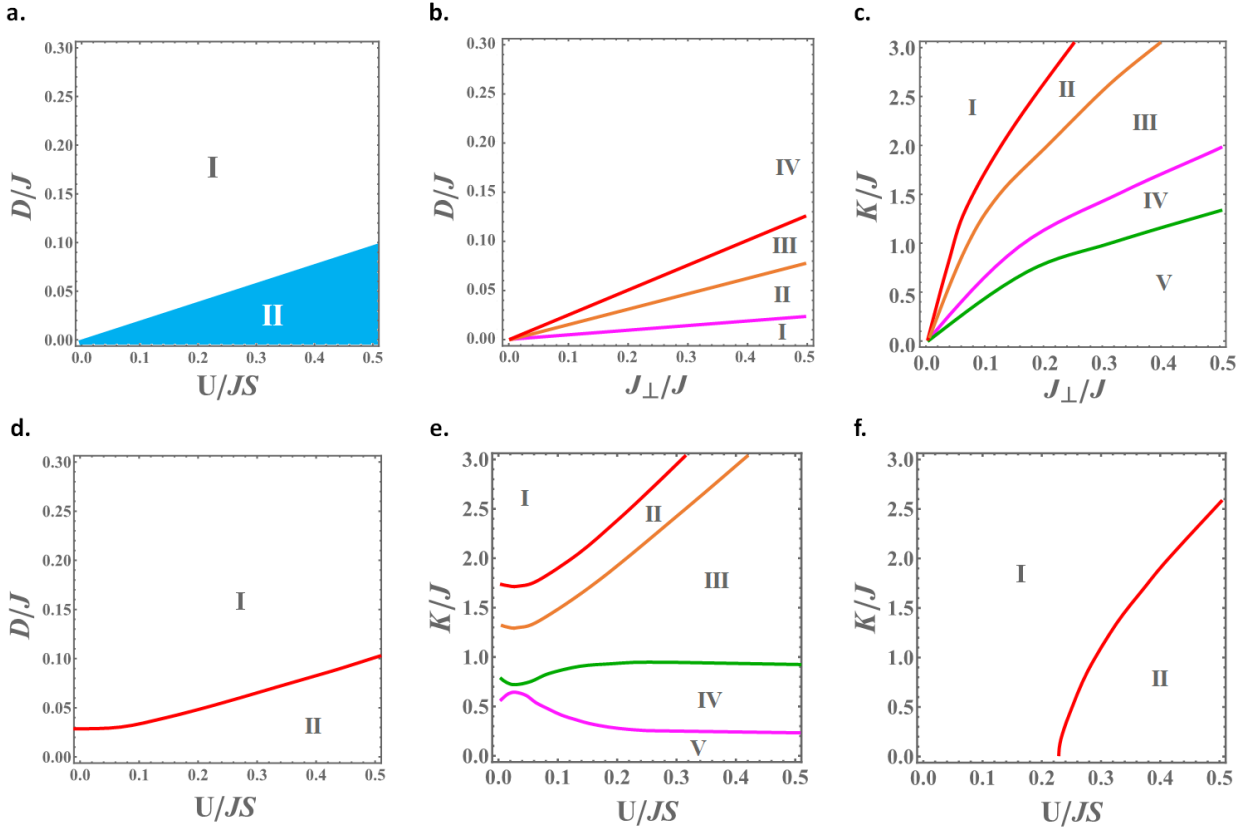

**Figure 5:** (a) TPD for the $D + U$ model in monoclinic LAFM bilayers with $J_\perp = -0.1J$. Phase I is gapped and topological, while phase II is ungapped with UNPs. (b) TPD for the $D$ model in $R_{33}$ stacked FM bilayers. (c) TPD for the $K$ model in $R_{33}$ stacked FM bilayers. (d), (e), and (f) are TPDs for $R_{33}$ stacked FM bilayers in the $D + U$, $K + U$, and $K + U + D$ models, respectively. These are calculated with $J_\perp = -0.1J$.

## III. Discussion

The stacking-dependent magnetism, SOCs, and electric control of magnetic order in 2D bilayer magnets have recently received exceptional attention given their potentials in spintronics. Nevertheless, the implications of these degrees of freedom on the magnon topology remain widely unexplored. The present study analyzed the magnonic topology and valley-polarization in 39 honeycomb bilayer models that span all relevant combinations of SOCs, magnetic configurations, stacking orders, and inversion symmetry breaking. The study is kept general, presenting numerical results over extensive ranges of the fundamental parameters to cover the broadest spectrum of vdWs magnets. The main conclusions are summarized in Table 3, revealing the remarkable topological dependence on all degrees of freedom included in our study.

| Model | FM Ground State | | |
|---|---|---|---|
| | **AB** | **Monoclinic** | $R_{33}$ |
| $U$ | • Topologically trivial with valley Hall conductivity. | • Topologically trivial with nonzero valley Hall conductivity. | • Topologically trivial with valley Hall conductivity. |
| $D$ | • Single topological phase $[0,-2,0,2]$. | • Single topological phase $[1,1,-1,-1]$. | • Topological phases: $[1,1,-1,-1]$, $[1,0,0,-1]$, and $[1,-1,1,-1]$.<br>• One trivial phase with standard or valley Hall conductivity depending on the value of $D$. |
| $K$ | • Single topological phase $[0,-2,0,2]$. | • Single topological phase $[1,1,-1,-1]$. | • Topological phases: $[1,1,-1,-1]$, $[1,0,0,-1]$, $[1,-1,1,-1]$, and $[1,-1,0,0]$.<br>• One trivial phase with standard or valley Hall conductivity depending on the value of $K$. |
| $K+D$ | • Single topological phase $[0,-2,0,2]$. | • Single topological phase $[1,1,-1,-1]$. | • Single topological phase $[1,1,-1,-1]$. |
| $D+U$ | • Topological phases: $[0,-2,0,2]$, $[0,-2,1,1]$, $[-1,-1,1,1]$, $[-1,1,-1,1]$, and $[0,0,-1,1]$.<br>• One trivial phase with standard or valley Hall conductivity depending on the value of $D$. | • Single topological phase $[1,1,-1,-1]$. | • Topological phases: $[1,1,-1,-1]$ and $[1,-1,1,-1]$. |
| $K+U$ | • Topological phases: $[0,-2,0,2]$, $[0,-2,1,1]$, $[-1,-1,1,1]$, $[-1,1,-1,1]$, and $[0,0,-1,1]$.<br>• One trivial phase with standard or valley Hall conductivity depending on the value of $K$. | • Single topological phase $[1,1,-1,-1]$. | • Topological phases: $[1,1,-1,-1]$, $[1,0,0,-1]$, $[1,-1,1,-1]$, and $[1,-1,0,0]$.<br>• One trivial phase with standard or valley Hall conductivity depending on the value of $K$. |
| $K+U+D$ | • Topological phases: $[0,-2,0,2]$, $[0,-2,1,1]$, $[-1,-1,1,1]$, $[-1,1,-1,1]$, and $[0,0,-1,1]$. | • Single topological phase $[1,1,-1,-1]$. | • Topological phases: $[1,1,-1,-1]$, $[1,0,0,-1]$, and $[1,-1,1,-1]$. |

**Table 3.** Main conclusions on the TPDs and magnon valley-polarization in FM bilayer models.

| Model | LAFM Ground State | | |
|---|---|---|---|
| | **AB** | **Monoclinic** | $R_{33}$ |
| $U$ | • Topologically trivial with valley Hall conductivity. | • The bands are degenerate. Does not support valley Hall transport. | • The bands are degenerate. Does not support valley Hall transport. |
| $D$ | • Topologically trivial. | • Topologically trivial. | • Topologically trivial. |
| $K$ | • Topologically trivial. | • Topologically trivial. | • Topologically trivial. |
| $D + U$ | • Single topological phase $[-1,1,1-1]$.<br>• One trivial phase with standard Hall conductivity. Valley Hall transport is unlikely. | • Single topological phase $[-1,1,1-1]$. | • Single topological phase $[-1,1,1-1]$. |
| $K + U$ | • Topological phases: $[-1,1,1-1]$, $[-1,2,0,-1]$, $[-1,-1,3,-1]$, $[0,-2,3,-1]$, $[0,-2,2,0]$, and $[0,1,-1,0]$.<br>• One trivial phase with standard or valley Hall conductivity depending on the value of $K$. | • Single topological phase $[-1,1,1-1]$. | • Single topological phase $[-1,1,1-1]$. |
| $K + U + D$ | • Topological phases: $[-1,1,1,-1]$ and $[-1,2,0,-1]$. | • Single topological phase $[-1,1,1-1]$. | • Single topological phase $[-1,1,1-1]$. |

**Table 4.** Main conclusions on the TPDs and magnon valley-polarization in LAFM bilayer models.

The work is further motivated by the open question regarding the nature of SOCs underlying 2D magnets[24]. We show that the standard Hall and Nernst conductivities carry signatures that can be matched with the type of SOC in several bilayer models. Indeed, magnetic materials' topological nature is best explored through their Hall and Nernst responses. Nevertheless, our study concludes that these conductivities are strongly related to the SOC's strength and bands' nonreciprocity, and they can survive in bilayers with zero Chern numbers.

Understanding the relation between bandgap closures and the structure of the TPDs is another primary objective. We revealed fundamental differences between the TPDs in FM and LAFM bilayers. The TPDs in FM bilayers present conventional bandgap closure manifolds associated with Dirac cones located at the $\pm K$ BZ corners. However, in LAFM bilayers, the TPDs are tailored primarily by the UNPs, introducing a novel bandgap closure scenario away from the high symmetry points. These contribute additional boundaries to the TPDs and enrich the topology.

We end with a brief discussion on the advantages of 2D magnetic bilayers over their monolayer version. Bilayers present additional degrees of freedom, detailed in our study, which can help tune their magnonic bands and topological response. Monolayers are restricted to a single topological phase in the presence of Kitaev and/or DM interactions. Consequently, monolayers' topology cannot be tuned, and the information extracted from their topological response is limited. Bilayers are hence superior from fundamental and applied perspectives.

## IV. Methods

The Berry curvatures and Chern numbers are determined using the approach developed in reference[79]. We first discretize the BZ with steps $\delta p$ in the $x$ and $y$ directions and numerically calculate the phase variations generated by infinitesimal displacements in the BZ,

$$U_i^x(\vec{p}) = \langle \epsilon_i(\vec{p} + \delta p\, \hat{p}_x) | \epsilon_i(\vec{p}) \rangle$$
$$U_i^y(\vec{p}) = \langle \epsilon_i(\vec{p} + \delta p\, \hat{p}_y) | \epsilon_i(\vec{p}) \rangle$$

(2)

Here, $|\epsilon_i\rangle$ denotes the eigenstate for the band $i$. The Wilson loop $W_i(\vec{p})$ can then be deduced as

$$W_i(\vec{p}) = U_i^x(\vec{p}) U_i^y(\vec{p} + \delta p\, \hat{p}_x) U_i^{x*}(\vec{p} + \delta p\, \hat{p}_y) U_i^{y*}(\vec{p})$$

(3)

The argument ($arg$-function) of $W_i(\vec{p})$ consequently yields the Berry curvatures $\mathcal{B}_i(\vec{p})$ via the identity

$$\mathcal{B}_i(\vec{p}) = \frac{\arg W_i(\vec{p})}{\delta p^2}$$

(4)

Knowing the Berry curvatures, the Chern numbers are determined through numerical integration over the BZ,

$$C_i = \frac{1}{2\pi} \iint_{BZ} \mathcal{B}_i(\vec{p}) dp_x dp_y$$

(5)

Topological Hall ($\kappa_{xy}$) and Nernst ($\alpha_{xy}$) conductivities can also be deduced from the Berry curvatures using the standard equations,

$$\kappa_{xy} = -\frac{k_B^2 T}{\hbar V} \sum_{\vec{p},i} c_2\left(g(\epsilon_i(\vec{p}))\right) \mathcal{B}_i(\vec{p})$$

(6)

$$\alpha_{xy} = \frac{k_B}{V} \sum_{\vec{p},i} c_1\left(g(\epsilon_i(\vec{p}))\right) \mathcal{B}_i(\vec{p})$$

(7)

with momentum $\vec{p}$ summed over the discretized BZ. The symbols $V$ and $k_B$ denote the volume of the system and the Boltzmann constant, respectively. We note the use of $k_B = \hbar = 1$ in our numerical calculation. The function $g(\epsilon_i) = \left[e^{\epsilon_i/k_B T} - 1\right]^{-1}$ stands for the Bose-Einstein distribution, $c_1(x) = (1+x)\ln(1+x) - x\ln x$, and $c_2(x) = (1+x)\left[\ln\left(\frac{1+x}{x}\right)\right]^2 - (\ln x)^2 - 2\text{Li}_2(-x)$. $\text{Li}_2$ is the dilogarithm function.

# Insights on magnon topology and valley-polarization in 2D bilayer quantum magnets

## Supplementary Information

D. Ghader

College of Engineering and Technology, American University of the Middle East, Egaila, Kuwait

**Supplementary Note 1: Derivation of the FM momentum-space Hamiltonians.**

We recall the expression for the real-space Hamiltonian

$$\mathcal{H} = -J \sum_{l,i,j} \vec{S}_{li} \cdot \vec{S}_{lj} - K \sum_{l,i,j} (\vec{S}_{li} \cdot \hat{\gamma}_{ij})(\vec{S}_{lj} \cdot \hat{\gamma}_{ij}) - J_\perp \sum_{i,j} \vec{S}_{1i} \cdot \vec{S}_{2j} + \sum_{l,m,n} D_{mn} \vec{S}_{lm} \cdot \vec{S}_{ln}^D$$
$$- \sum_{l,i} U_l \hat{z} \cdot \vec{S}_{li} - \mathcal{A} \sum_{l,i} (\vec{S}_{li} \cdot \hat{z})^2$$

(1)

In the semi-classical linear spin-wave approach, the spins are treated as numerical vectors, and the spin dynamics are derived from the Landau-Lifshitz equations of motion. The Holstein-Primakov method yields identical results. For FM bilayers, $\mathcal{H}$ can be rewritten in the form $\mathcal{H} = \sum_{l,i} \vec{S}_{li} \cdot \vec{H}_{li}^{FM}$, which introduces the effective magnetic fields $\vec{H}_{li}^{FM}$ acting on spins $\vec{S}_{li}$. Employing Fourier transformation, the fields corresponding to the four sublattices $\{A_1, B_1, A_2, B_2\}$ in the AB-stacked bilayers reduce to

$$\vec{H}_{A_1}^{FM}(\vec{p}) = \left[-\mathcal{J}_{xx} u_x^{B_1} - i\mathcal{J}_{xy} u_y^{B_1} + if_D u_y^{A_1}\right] \hat{x} +$$
$$\left[-\mathcal{J}_{yy} u_y^{B_1} - i\mathcal{J}_{xy} u_x^{B_1} - if_D u_x^{A_1}\right] \hat{y} +$$
$$\left[-(3J + 2\mathcal{A} + K + U)S - \mathcal{J}_{zx} u_x^{B_1} - i\sqrt{2}\,\mathcal{J}_{xy} u_y^{B_1}\right] \hat{z}$$

(2a)

$$\vec{H}_{B_1}^{FM}(\vec{p}) = \left[-\mathcal{J}_{xx}^* u_x^{A_1} + i\mathcal{J}_{xy}^* u_y^{A_1} - J_\perp S u_x^{A_2} - if_D u_y^{B_1}\right] \hat{x} +$$
$$\left[-\mathcal{J}_{yy}^* u_y^{A_1} + i\mathcal{J}_{xy}^* u_x^{A_1} - J_\perp S u_y^{A_2} + if_D u_x^{B_1}\right] \hat{y} +$$
$$\left[-(3J + 2\mathcal{A} + J_\perp + K + U)S - \mathcal{J}_{zx}^* u_x^{A_1} + i\sqrt{2}\,\mathcal{J}_{xy}^* u_y^{A_1}\right] \hat{z}$$

(2b)

$$\vec{H}_{A_2}^{FM}(\vec{p}) = \left[-\mathcal{J}_{xx}u_x^{B_2} - i\mathcal{J}_{xy}u_y^{B_2} - J_\perp S u_x^{B_1} + if_D u_y^{A_2}\right]\hat{x} +$$

$$\left[-\mathcal{J}_{yy}u_y^{B_2} - i\mathcal{J}_{xy}u_x^{B_2} - J_\perp S u_y^{B_1} - if_D u_x^{A_2}\right]\hat{y} +$$

$$\left[-(3J + 2\mathcal{A} + J_\perp + K - U)S - \mathcal{J}_{zx}u_x^{B_2} - i\sqrt{2}\,\mathcal{J}_{xy}u_y^{B_2}\right]\hat{z}$$

(2c)

$$\vec{H}_{B_2}^{FM}(\vec{p}) = \left[-\mathcal{J}_{xx}^* u_x^{A_2} + i\mathcal{J}_{xy}^* u_y^{A_2} - if_D u_y^{B_2}\right]\hat{x} +$$

$$\left[-\mathcal{J}_{yy}^* u_y^{A_2} + i\mathcal{J}_{xy}^* u_x^{A_2} + if_D u_x^{B_2}\right]\hat{y} +$$

$$\left[-(3J + 2\mathcal{A} + K - U)S - \mathcal{J}_{zx}^* u_x^{A_2} + i\sqrt{2}\,\mathcal{J}_{xy}^* u_y^{A_2}\right]\hat{z}$$

(2d)

with $\mathcal{J}_{xx} = S\left(\frac{2}{3}K + J\right)f_1(\vec{p}) + S\left(\frac{1}{3}K + 2J\right)f_2(\vec{p})$, $\mathcal{J}_{xy} = \frac{1}{\sqrt{3}}KSf_3(\vec{p})$, $\mathcal{J}_{yy} = JSf_1(\vec{p}) + S(K + 2J)f_2(\vec{p})$, $\mathcal{J}_{zx} = \frac{\sqrt{2}}{3}KS[-f_1(\vec{p}) + f_2(\vec{p})]$, $f_1(\vec{p}) = e^{i\frac{1}{\sqrt{3}}p_y}$, $f_2(\vec{p}) = e^{-i\frac{1}{2\sqrt{3}}p_y}\cos\left(\frac{p_x}{2}\right)$, $f_3(\vec{p}) = e^{-i\frac{1}{2\sqrt{3}}p_y}\sin\left(\frac{p_x}{2}\right)$, and $f_D(\vec{p}) = 4DS\sin\left(\frac{p_x}{2}\right)\cos\left(\frac{\sqrt{3}p_y}{2}\right) - 2DS\sin(p_x)$. $\vec{p}$ denotes momentum, while $S$ and $u_\beta^\alpha(\vec{p})$ represent the constant z-component of the spin and the Fourier transform coefficients, respectively.

The momentum-space Hamiltonian $\mathcal{H}_{FM}^{AB}(\vec{p})$ can now be derived from the Landau-Lifshitz equations, $\partial_t \vec{S} = \vec{S} \times \vec{H}^{FM}$, generating eight coupled scalar equations in the coefficients $u^{\alpha_l} = u_x^{\alpha_l} + iu_y^{\alpha_l}$ and $u^{\alpha_l *} = u_x^{\alpha_l} - iu_y^{\alpha_l}$ ($\alpha = A, B$). The derived equations can be grouped in the form of an eigenenergy matrix equation $[\mathcal{H}_{FM}^{AB}(\vec{p}) - \epsilon(\vec{p})I]|u(\vec{p})\rangle = 0$, with the $8 \times 8$ momentum-space Hamiltonian given explicitly as

$$\mathcal{H}_{FM}^{AB}(\vec{p}) = \begin{pmatrix} \mathcal{F}_{11}^{AB} & \mathcal{F}_{12}^{AB} \\ \mathcal{F}_{21}^{AB} & \mathcal{F}_{22}^{AB} \end{pmatrix},$$

with

$$\mathcal{F}_{11}^{AB} = \begin{pmatrix} c_1 + f_D + U & -c_2(f_1 + 2f_2) & 0 & c_3(\sqrt{3}f_3 - f_1 + f_2) \\ -c_2(f_1^* + 2f_2^*) & c_1 + J_\perp S - f_D + U & c_3(-\sqrt{3}f_3^* - f_1^* + f_2^*) & 0 \\ 0 & c_3(\sqrt{3}f_3 + f_1 - f_2) & -c_1 + f_D - U & c_2(f_1 + 2f_2) \\ c_3(-\sqrt{3}f_3^* + f_1^* - f_2^*) & 0 & c_2(f_1^* + 2f_2^*) & -c_1 - J_\perp S - f_D - U \end{pmatrix},$$

$$\mathcal{F}_{22}^{AB} = \begin{pmatrix} c_1 + J_\perp S + f_D - U & -c_2(f_1 + 2f_2) & 0 & c_3(\sqrt{3}f_3 - f_1 + f_2) \\ -c_2(f_1^* + 2f_2^*) & c_1 - f_D - U & c_3(-\sqrt{3}f_3^* - f_1^* + f_2^*) & 0 \\ 0 & c_3(\sqrt{3}f_3 + f_1 - f_2) & -c_1 - J_\perp S + f_D + U & c_2(f_1 + 2f_2) \\ c_3(-\sqrt{3}f_3^* + f_1^* - f_2^*) & 0 & c_2(f_1^* + 2f_2^*) & -c_1 - f_D + U \end{pmatrix},$$

$$\mathcal{F}_{12}^{AB} = J_\perp S \begin{pmatrix} 0 & -1 & 0 & 0 \\ 0 & 0 & 0 & 0 \\ 0 & 0 & 0 & 1 \\ 0 & 0 & 0 & 0 \end{pmatrix}, \quad \mathcal{F}_{21}^{AB} = J_\perp S \begin{pmatrix} 0 & 0 & 0 & 0 \\ -1 & 0 & 0 & 0 \\ 0 & 0 & 0 & 0 \\ 0 & 0 & 1 & 0 \end{pmatrix}.$$

The constants $c_1 = (3J + 2\mathcal{A} + K)S$, $c_2 = \left(\frac{1}{3}K + J\right)S$, and $c_3 = \frac{1}{3}KS$.

The same steps are employed to derive the momentum-space Hamiltonians in the monoclinic case. In monoclinic bilayers, every site interacts with two interlayer nearest neighbors (Supplementary Fig.1a). The corresponding Hamiltonian becomes

$$\mathcal{H}_{FM}^{Mono}(\vec{p}) = \begin{pmatrix} \mathcal{F}_{11}^{Mono} & \mathcal{F}_{12}^{Mono} \\ \mathcal{F}_{21}^{Mono} & \mathcal{F}_{22}^{Mono} \end{pmatrix},$$

with

$$\mathcal{F}_{11}^{Mono} = \begin{pmatrix} c_1 + f_D + U & -c_2(f_1 + 2f_2) & 0 & c_3(\sqrt{3}f_3 - f_1 + f_2) \\ -c_2(f_1^* + 2f_2^*) & c_1 + 2J_\perp S - f_D + U & c_3(-\sqrt{3}f_3^* - f_1^* + f_2^*) & 0 \\ 0 & c_3(\sqrt{3}f_3 + f_1 - f_2) & -c_1 + f_D - U & c_2(f_1 + 2f_2) \\ c_3(-\sqrt{3}f_3^* + f_1^* - f_2^*) & 0 & c_2(f_1^* + 2f_2^*) & -c_1 - 2J_\perp S - f_D - U \end{pmatrix},$$

$$\mathcal{F}_{22}^{Mono} = \begin{pmatrix} c_1 + 2J_\perp S + f_D - U & -c_2(f_1 + 2f_2) & 0 & c_3(\sqrt{3}f_3 - f_1 + f_2) \\ -c_2(f_1^* + 2f_2^*) & c_1 - f_D - U & c_3(-\sqrt{3}f_3^* - f_1^* + f_2^*) & 0 \\ 0 & c_3(\sqrt{3}f_3 + f_1 - f_2) & -c_1 - 2J_\perp S + f_D + U & c_2(f_1 + 2f_2) \\ c_3(-\sqrt{3}f_3^* + f_1^* - f_2^*) & 0 & c_2(f_1^* + 2f_2^*) & -c_1 - f_D + U \end{pmatrix},$$

$$\mathcal{F}_{12}^{Mono} = J_\perp S \begin{pmatrix} -e^{\frac{ip_x}{3}} & -e^{-\frac{ip_x}{6}} e^{-\frac{ip_y}{2\sqrt{3}}} & 0 & 0 \\ -e^{-\frac{ip_x}{6}} e^{\frac{ip_y}{2\sqrt{3}}} & -e^{\frac{ip_x}{3}} & 0 & 0 \\ 0 & 0 & e^{\frac{ip_x}{3}} & e^{-\frac{ip_x}{6}} e^{-\frac{ip_y}{2\sqrt{3}}} \\ 0 & 0 & e^{-\frac{ip_x}{6}} e^{\frac{ip_y}{2\sqrt{3}}} & e^{\frac{ip_x}{3}} \end{pmatrix},$$

$$\mathcal{F}_{21}^{Mono} = J_\perp S \begin{pmatrix} -e^{-\frac{ip_x}{3}} & -e^{\frac{ip_x}{6}}e^{-\frac{ip_y}{2\sqrt{3}}} & 0 & 0 \\ -e^{\frac{ip_x}{6}}e^{\frac{ip_y}{2\sqrt{3}}} & -e^{-\frac{ip_x}{3}} & 0 & 0 \\ 0 & 0 & e^{-\frac{ip_x}{3}} & e^{\frac{ip_x}{6}}e^{-\frac{ip_y}{2\sqrt{3}}} \\ 0 & 0 & e^{\frac{ip_x}{6}}e^{\frac{ip_y}{2\sqrt{3}}} & e^{-\frac{ip_x}{3}} \end{pmatrix}.$$

The $R_{33}$ configuration is illustrated in Supplementary Fig.1b, which modifies the Hamiltonian as

$$\mathcal{H}_{FM}^R(\vec{p}) = \begin{pmatrix} \mathcal{F}_{11}^R & \mathcal{F}_{12}^R \\ \mathcal{F}_{21}^R & \mathcal{F}_{22}^R \end{pmatrix},$$

with

$$\mathcal{F}_{11}^R = \mathcal{F}_{11}^{AB}, \quad \mathcal{F}_{22}^R = \mathcal{F}_{22}^{AB}, \quad \mathcal{F}_{12}^R = \mathcal{F}_{21}^R = J_\perp S \begin{pmatrix} -e^{-\frac{ip_y}{2\sqrt{3}}} & 0 & 0 & 0 \\ 0 & -e^{\frac{ip_y}{2\sqrt{3}}} & 0 & 0 \\ 0 & 0 & e^{-\frac{ip_y}{2\sqrt{3}}} & 0 \\ 0 & 0 & 0 & e^{\frac{ip_y}{2\sqrt{3}}} \end{pmatrix}.$$

**Supplementary Note 2: Derivation of the LAFM momentum-space Hamiltonians.**

In LAFM bilayers, the spins in layer 2 are expressed as $\vec{S} = (-S_x, S_y, -S)$. The LAFM effective fields in the AB-stacking read

$$\vec{H}_{A_1}^{AFM}(\vec{p}, t) = \vec{H}_{A_1}^{FM}(\vec{p}, t)$$

(3a)

$$\vec{H}_{B_1}^{AFM}(\vec{p}, t) = \left[ -\mathcal{J}_{xx}^* u_x^{A_1} + i\mathcal{J}_{xy}^* u_y^{A_1} + J_\perp S u_x^{A_2} - if_D u_y^{B_1} \right] \hat{x} +$$

$$\left[ -\mathcal{J}_{yy}^* u_y^{A_1} + i\mathcal{J}_{xy}^* u_x^{A_1} - J_\perp S u_y^{A_2} + if_D u_x^{B_1} \right] \hat{y} +$$

$$\left[ -(3J + 2\mathcal{A} - J_\perp + K + U)S - \mathcal{J}_{zx}^* u_x^{A_1} + i\sqrt{2}\, \mathcal{J}_{xy}^* u_y^{A_1} \right] \hat{z}$$

(3b)

$$\vec{H}^{AFM}_{A_2}(\vec{p},t) = \left[\mathcal{J}_{xx}u_x^{B_2} - i\mathcal{J}_{xy}u_y^{B_2} - J_\perp S u_x^{B_1} + if_D u_y^{A_2}\right]\hat{x} +$$

$$\left[-\mathcal{J}_{yy}u_y^{B_2} + i\mathcal{J}_{xy}u_x^{B_2} - J_\perp S u_y^{B_1} + if_D u_x^{A_2}\right]\hat{y} +$$

$$\left[(3J + 2\mathcal{A} - J_\perp + K - U)S + \mathcal{J}_{zx}u_x^{B_2} - i\sqrt{2}\,\mathcal{J}_{xy}u_y^{B_2}\right]\hat{z}$$

(3c)

$$\vec{H}^{AFM}_{B_2}(\vec{p},t) = \left[\mathcal{J}^*_{xx}u_x^{A_2} + i\mathcal{J}^*_{xy}u_y^{A_2} - if_D u_y^{B_2}\right]\hat{x} +$$

$$\left[-\mathcal{J}^*_{yy}u_y^{A_2} - i\mathcal{J}^*_{xy}u_x^{A_2} - if_D u_x^{B_2}\right]\hat{y} +$$

$$\left[(3J + 2\mathcal{A} + K - U)S + \mathcal{J}^*_{zx}u_x^{A_2} + i\sqrt{2}\,\mathcal{J}^*_{xy}u_y^{A_2}\right]\hat{z}$$

(3d)

The Landau-Lifshitz equations yield the LAFM momentum-space Hamiltonian for AB-stacking as

$$\mathcal{H}^{AB}_{AFM}(\vec{p}) = \begin{pmatrix} \mathcal{G}^{AB}_{11} & \mathcal{G}^{AB}_{12} \\ \mathcal{G}^{AB}_{21} & \mathcal{G}^{AB}_{22} \end{pmatrix}$$

with

$$\mathcal{G}^{AB}_{11} = \begin{pmatrix} c_1 + f_D + U & -c_2(f_1 + 2f_2) & 0 & c_3(\sqrt{3}f_3 - f_1 + f_2) \\ -c_2(f_1^* + 2f_2^*) & c_1 - J_\perp S - f_D + U & c_3(-\sqrt{3}f_3^* - f_1^* + f_2^*) & 0 \\ 0 & c_3(\sqrt{3}f_3 + f_1 - f_2) & -c_1 + f_D - U & c_2(f_1 + 2f_2) \\ c_3(-\sqrt{3}f_3^* + f_1^* - f_2^*) & 0 & c_2(f_1^* + 2f_2^*) & -c_1 + J_\perp S - f_D - U \end{pmatrix},$$

$$\mathcal{G}^{AB}_{22} = \begin{pmatrix} c_1 - J_\perp S - f_D - U & -c_2(f_1 + 2f_2) & 0 & c_3(-\sqrt{3}f_3 - f_1 + f_2) \\ -c_2(f_1^* + 2f_2^*) & c_1 + f_D - U & c_3(\sqrt{3}f_3^* - f_1^* + f_2^*) & 0 \\ 0 & c_3(-\sqrt{3}f_3 + f_1 - f_2) & -c_1 + J_\perp S - f_D + U & c_2(f_1 + 2f_2) \\ c_3(\sqrt{3}f_3^* + f_1^* - f_2^*) & 0 & c_2(f_1^* + 2f_2^*) & -c_1 + f_D + U \end{pmatrix},$$

$$\mathcal{G}^{AB}_{12} = \begin{pmatrix} 0 & 0 & 0 & 0 \\ 0 & 0 & J_\perp S & 0 \\ 0 & 0 & 0 & 0 \\ -J_\perp S & 0 & 0 & 0 \end{pmatrix}, \text{ and } \mathcal{G}^{AB}_{21} = \begin{pmatrix} 0 & 0 & 0 & J_\perp S \\ 0 & 0 & 0 & 0 \\ 0 & -J_\perp S & 0 & 0 \\ 0 & 0 & 0 & 0 \end{pmatrix}.$$

For the remaining stacking configurations, we have

$$\mathcal{H}_{AFM}^{Mono}(\vec{p}) = \begin{pmatrix} \mathcal{G}_{11}^{Mono} & \mathcal{G}_{12}^{Mono} \\ \mathcal{G}_{21}^{Mono} & \mathcal{G}_{22}^{Mono} \end{pmatrix}$$

with

$$\mathcal{G}_{11}^{Mono} = \begin{pmatrix} c_1 + f_D + U & -c_2(f_1 + 2f_2) & 0 & c_3(\sqrt{3}f_3 - f_1 + f_2) \\ -c_2(f_1^* + 2f_2^*) & c_1 - 2J_\perp S - f_D + U & c_3(-\sqrt{3}f_3^* - f_1^* + f_2^*) & 0 \\ 0 & c_3(\sqrt{3}f_3 + f_1 - f_2) & -c_1 + f_D - U & c_2(f_1 + 2f_2) \\ c_3(-\sqrt{3}f_3^* + f_1^* - f_2^*) & 0 & c_2(f_1^* + 2f_2^*) & -c_1 + 2J_\perp S - f_D - U \end{pmatrix},$$

$$\mathcal{G}_{22}^{Mono} = \begin{pmatrix} c_1 - 2J_\perp S - f_D - U & -c_2(f_1 + 2f_2) & 0 & c_3(-\sqrt{3}f_3 - f_1 + f_2) \\ -c_2(f_1^* + 2f_2^*) & c_1 + f_D - U & c_3(\sqrt{3}f_3^* - f_1^* + f_2^*) & 0 \\ 0 & c_3(-\sqrt{3}f_3 + f_1 - f_2) & -c_1 + 2J_\perp S - f_D + U & c_2(f_1 + 2f_2) \\ c_3(\sqrt{3}f_3^* + f_1^* - f_2^*) & 0 & c_2(f_1^* + 2f_2^*) & -c_1 + f_D + U \end{pmatrix},$$

$$\mathcal{G}_{12}^{Mono} = J_\perp S \begin{pmatrix} 0 & 0 & e^{\frac{ip_x}{3}} & e^{-\frac{ip_x}{6}} e^{-\frac{ip_y}{2\sqrt{3}}} \\ 0 & 0 & e^{-\frac{ip_x}{6}} e^{\frac{ip_y}{2\sqrt{3}}} & e^{\frac{ip_x}{3}} \\ -e^{\frac{ip_x}{3}} & -e^{-\frac{ip_x}{6}} e^{-\frac{ip_y}{2\sqrt{3}}} & 0 & 0 \\ -e^{-\frac{ip_x}{6}} e^{\frac{ip_y}{2\sqrt{3}}} & -e^{\frac{ip_x}{3}} & 0 & 0 \end{pmatrix},$$

$$\mathcal{G}_{21}^{Mono} = J_\perp S \begin{pmatrix} 0 & 0 & e^{-\frac{ip_x}{3}} & e^{\frac{ip_x}{6}} e^{-\frac{ip_y}{2\sqrt{3}}} \\ 0 & 0 & e^{\frac{ip_x}{6}} e^{\frac{ip_y}{2\sqrt{3}}} & e^{-\frac{ip_x}{3}} \\ -e^{-\frac{ip_x}{3}} & -e^{\frac{ip_x}{6}} e^{-\frac{ip_y}{2\sqrt{3}}} & 0 & 0 \\ -e^{\frac{ip_x}{6}} e^{\frac{ip_y}{2\sqrt{3}}} & -e^{-\frac{ip_x}{3}} & 0 & 0 \end{pmatrix},$$

while

$$\mathcal{H}_{AFM}^{R}(\vec{p}) = \begin{pmatrix} \mathcal{G}_{11}^{R} & \mathcal{G}_{12}^{R} \\ \mathcal{G}_{21}^{R} & \mathcal{G}_{22}^{R} \end{pmatrix}$$

with $\mathcal{G}_{11}^{R} = \mathcal{G}_{11}^{AB}$, $\mathcal{G}_{22}^{R} = \mathcal{G}_{22}^{AB}$, and $\mathcal{G}_{12}^{R} = \mathcal{G}_{21}^{R} = J_\perp S \begin{pmatrix} 0 & 0 & 0 & e^{-\frac{ip_y}{2\sqrt{3}}} \\ 0 & 0 & e^{\frac{ip_y}{2\sqrt{3}}} & 0 \\ 0 & -e^{-\frac{ip_y}{2\sqrt{3}}} & 0 & 0 \\ -e^{\frac{ip_y}{2\sqrt{3}}} & 0 & 0 & 0 \end{pmatrix}$.

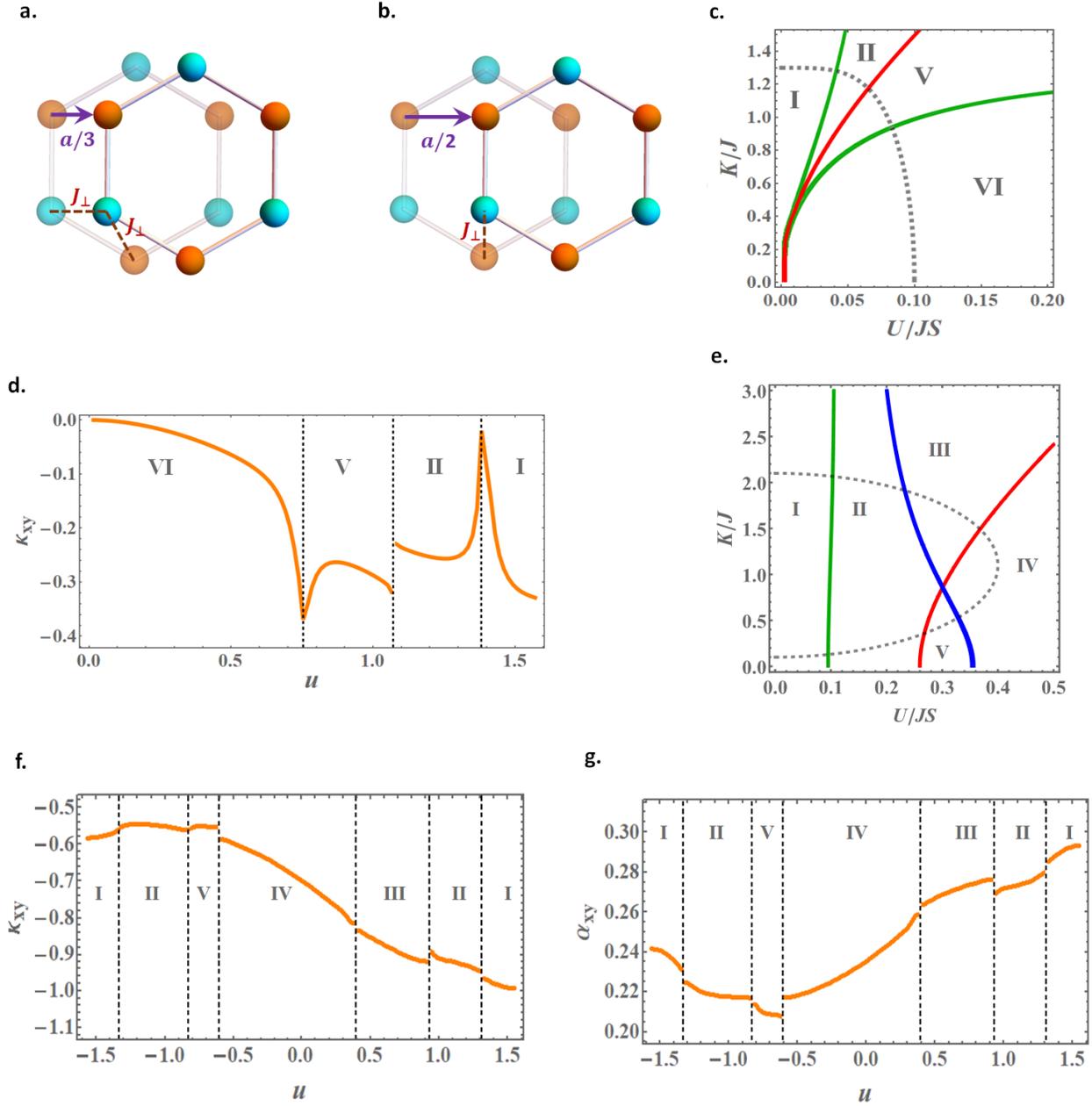

**Supplementary Figure 1:** Schematic representation for the (a) monoclinic and (b) $R_{33}$ stacking. (c) TPD for the $K + U$ model with $J_\perp = 0.2J$ and weak ED. (d) Nonadiabatic evolution of the Hall conductivity along the nonadiabatic path in (c). (e) TPD in the AB-stacked FM $K + U + D$ model with $D = 0.05J$ and $J_\perp = 0.3J$. (f) and (g) illustrate the nonadiabatic evolution of the Hall and Nernst conductivities, respectively. These are plotted along the nonadiabatic path in (b).